# Quantitative Microscale Thermometry in Droplets Loaded with Gold Nanoparticles

*Lucas Sixdenier,[a],\* Guillaume Baffou,[b] Christophe Tribet,[a] Emmanuelle Marie [a],\**

[a] PASTEUR, Département de chimie, École normale supérieure, PSL University, Sorbonne Université, CNRS, 75005 Paris, France

[b] Institut Fresnel, CNRS, Aix Marseille University, Centrale Marseille, Marseille, France

ABSTRACT

Gold nanoparticles (AuNPs) are increasingly used for their thermoplasmonic properties, i.e. their ability to convert light into heat upon plasmon resonance. However, measuring temperature gradients generated at the microscale by assemblies of AuNPs remains challenging, especially when they are randomly distributed in three dimensions. Here, we introduce a label-free thermometry approach, combining optical wavefront microscopy and numerical simulations, to infer the heating power dissipated by a three-dimensional model system consisting in emulsion microdroplets loaded with AuNPs. This approach gives access to the temperature reached in the core of droplets upon irradiation without need of extrinsic calibration. These quantitative results are validated qualitatively *via* the observation of the phase transition of a thermoresponsive polymer added in the droplet as an *in situ* thermal probe. This versatile thermometry approach is promising for non-invasive temperature measurements in various three-dimensional microsystems involving AuNPs as colloidal heat sources, such as several light-responsive drug delivery systems.

## Introduction

When localized in gold nanoparticles (AuNPs), light-excited surface plasmons quickly dissipate energy under the form of heat, making AuNPs efficient light-to-heat converters at the nanoscale.[1,2] This so-called thermoplasmonic effect has the advantage to be spectrally tunable in the visible and near-IR range[3,4] by tailoring the size and/or morphology of the AuNPs through now well-controlled and easy-accessible synthesis pathways.[5,6] These unique properties have led to an increasing use of AuNPs as colloidal heat sources for diverse applications,[7,8] such as photothermal therapy,[9,10] light-induced molecular delivery,[11,12] cell manipulation,[13–15] heat-assisted chemistry,[16–18] or 3D printing.[19] Since thermoplasmonic effects are highly robust and manifest as soon as AuNPs are irradiated at the appropriate wavelength, generating heat with AuNPs is not challenging *per se*. However, being able to precisely measure temperature gradients induced around AuNPs (and more generally around any plasmonic nanostructure) remains a major challenge since nano- and microscale temperature fields are obviously not accessible to standard (macroscale) thermometry techniques.[20]

So far, *in situ* quantification of thermoplasmonic heating has been essentially based on the fluorescence of extrinsic reporters dispersed in the vicinity of the AuNPs under study. Indeed, many fluorescence features are temperature-dependent such as intensity,[21–23] life-time,[24,25] spectrum,[26] or anisotropy.[27,28] Thus, measuring changes of these properties upon thermoplasmonic heating may enable the determination of local temperature values. However, using fluorescent reporters may not only affect the chemical environment of the system under study, but may also suffer from photo- or thermo-bleaching that would bias the temperature measurements.[29] As an alternative to fluorescence, quantitative phase imaging turned out to be relevant for temperature mapping, because measuring the phase of light is label-free, non-invasive, and highly sensitive to temperature variations.[30] Indeed, temperature gradients generated by thermoplasmonic sources upon light excitation result in variations of refractive index in the surrounding medium, which in turn imprint a phase delay, or equivalently a wavefront distortion, to the light used to image the sample. Hence, recording wavefront distortions upon sample imaging should – in principle – enable one to retrieve the temperature fields generated *in situ* by plasmonic nanostructures.

Among quantitative phase imaging techniques, cross-grating wavefront microscopy (CGM), also known as quadriwave lateral shearing interferometry in the literature, is a diffraction-based wavefront imaging technique that has several advantages: it benefits from high signal/noise ratio and high spatio-temporal resolution, and it is compact and very little sensitive to environmental perturbations.[31,32] CGM provides both intensity and wavefront images of a sample from the analysis of the interferogram produced by a $0$-$\pi$ checkerboard-like diffraction grating placed at a millimeter distance from the sensor of a camera.[31] CGM has been successfully used to measure temperature fields and heating power densities at the level of single AuNPs or arrays of AuNPs,[33,34] demonstrating the high potential of this technique for *in situ* nanoscale thermometry without resorting to any extrinsic calibration. However, converting the wavefront image into a temperature image with this quantitative approach relies on the possibility to define a thermal Green's function of the system. This requirement makes this technique suited for simple environment geometries, typically for nanostructures lying upon a flat interface between two semi-infinite media,[33–36] but not for more complex heat source distributions, especially three-dimensional ones.

In this paper, we demonstrate the capability of CGM to provide label-free *in situ* microscale temperature measurements in a three-dimensional thermoplasmonic system, by combining CGM measurements and simulation-based inference of temperature fields. We illustrate this approach on water droplets loaded with randomly dispersed AuNPs in their volume, and dispersed in an oil phase. This system constitutes a relevant model of complex three-dimensional thermoplasmonic source since (i) the calculation of a thermal Green's function for a random distribution of several thousands of AuNPs is essentially ruled out, and (ii) the geometry of the system (spherical microdroplets) is well defined and suitable for heat transfer numerical simulations. In addition, emulsion droplets loaded with AuNPs is a versatile system that has been recently used for droplet manipulation,[37] for advanced sensing,[38,39] or as precursors of light-responsive encapsulation-and-release systems.[40,41]

CGM wavefront images of AuNP-loaded droplets were recorded upon light excitation to extract the wavefront distortion signature of thermoplasmonic effects. In parallel, the wavefront signature of an isolated droplet and the associated temperature field were predicted numerically by solving the steady-state heat equation in the appropriate geometry. Eventually, the heating power dissipated *in situ* and the temperature effectively reached in the droplet core were inferred from a comparison between experimental and numerical data. This microscale thermometry approach was implemented on isolated droplets of different sizes, and was validated *in situ* by probing a qualitative thermosensitive phenomenon, namely the heating-induced collapse transition of a thermoresponsive polymer. To our knowledge, this work constitutes the first demonstration of label-free quantitative thermometry in a three-dimensional thermoplasmonic system.

## Results and discussion

The system under study is a water-in-fluorinated oil emulsion in which the water droplets were loaded with 50-nm gold nanoparticles (AuNPs) at a concentration of 0.5 g/L ($2.2 \times 10^{-3}$ vol%). To preserve the integrity of the droplets, especially against heat-induced destabilization, the water/oil interface was stabilized by interfacial complexation[42] between an oil-soluble anionic surfactant (namely Krytox), and a water-soluble cationic copolymer made of a poly(L-lysine) (PLL) backbone grafted with poly(N-isopropylacrylamide) (PNIPAM) strands[43] (see Materials and Methods in SI). PNIPAM is a thermoresponsive polymer exhibiting a lower critical solution temperature (LCST) in water, above which it switches from hydrophilic to hydrophobic and collapses into insoluble colloidal aggregates.[44] Interestingly, the poly(L-lysine)-g-poly(N-isopropylacrylamide) copolymer (PLL-g-PNIPAM) was endowed with the same thermoresponsive behavior,[43] exhibiting a LCST of 33 °C (**Figure S1** in SI). Thanks to this property, PLL-g-PNIPAM (used at a concentration of 20 g/L) was not only used as a partner for interfacial complexation, but also as a qualitative *in situ* temperature probe upon thermoplasmonic heating, as the appearance of polymer granules in the droplet core would indicate that the temperature has been raised above 33 °C.

For cross-grating wavefront microscopy (CGM) imaging experiments the emulsion was injected in a rectangle glass capillary (Materials and Methods in SI). Since water is less dense than the continuous oil phase, the droplets were lying onto the upper internal side of the capillary (**Figure S2** in SI). **Figure 1** shows intensity micrographs of a droplet uniformly irradiated with a laser beam operating at 532 nm (corresponding to the plasmon wavelength for spherical AuNPs). Thermoplasmonic heating of the droplet above 33 °C was indicated by the appearance of polymer granules in the droplet core above a certain threshold in laser irradiance (i.e. power per surface unit). Here, the collapse of PLL-g-PNIPAM plays the role of a qualitative "single-value" temperature probe, but does not enable the determination of the absolute temperature reached in the droplet's core for any laser irradiance.

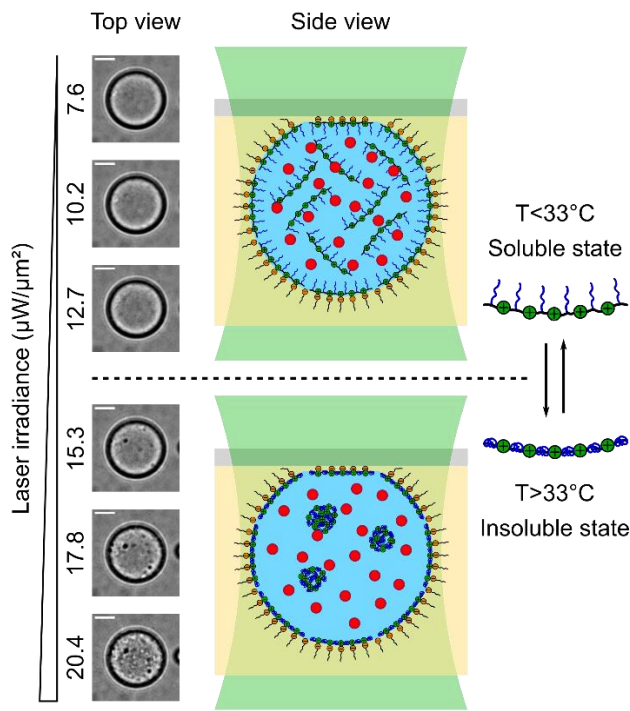

**Figure 1.** Qualitative single-value thermometry approach based on the collapse of thermoresponsive poly(L-lysine)-g-poly(N-isopropylacrylamide) polymer (PLL-g-PNIPAM) in a water droplet loaded with gold nanoparticles (red dots) upon plasmonic excitation. The droplet – in contact with the upper internal side of a capillary – was exposed to a 532-nm laser at increasing irradiance until the appearance of polymer granules in the droplet core. This phenomenon indicated that a temperature of 33 °C – corresponding to the LCST of PLL-g-PNIPAM – had been exceeded. PLL-g-PNIPAM was also used to stabilize the droplet interface *via* electrostatic complexation with an oil-soluble surfactant (Krytox) across the water/oil interface. Scale bar in the micrographs = 10 µm.

By quantitatively measuring wavefront distortions of the imaging light as it propagates through the sample, CGM maps the optical path difference (OPD, denoted $\delta\ell$) created by the imaged object. From a physical point of view, the OPD corresponds to the integral of the local variation of refractive index $n - n_e$ (where $n$ and $n_e$ are the indices of the object and its environment respectively) over the sample thickness, i.e. $\delta\ell = \int (n(z) - n_e)\mathrm{d}z$ – assuming that light propagates along the $z$ axis (this assumption being essentially valid in the oil phase, and for light passing through the upper cap of the droplets). In this work, all the OPD values measured were negative (as heating induces a decrease of refractive indices), but have been converted into positive values in data treatment and figures for convenience.

In the case of the sample described in Figure 1, the origin of variations of refractive index is two-fold: (i) a mismatch of refractive index intrinsically exists between the sample (here the droplet) and its environment (here the oil phase), and (ii) thermoplasmonic heating induces variations of refractive indices because of their temperature dependence (quantified by a $(\mathrm{d}n/\mathrm{d}T)$ coefficient). Only the second contribution was of interest for our study. To specifically isolate it, a first OPD image was acquired without heating (i.e. before turning the laser on), and was subsequently subtracted to the OPD images acquired upon laser excitation (**Figure 2**).

Since OPD images are generated by integration of wavefront gradients up to a constant, they come along with an arbitrary offset. Ideally, this offset could be determined by measuring the OPD infinitely far from the droplet (where there is no temperature increase) and setting it as the origin of OPD ($\delta\ell = 0$). However, due to the limited field of view of the microscope and the long-range temperature profile (slowly varying as $1/r$), we could not capture a large enough image to easily determine the proper OPD origin. Consequently, an offset was *arbitrarily* set to the experimental OPD values so that it equals zero at $\sim 50$ µm from the droplet center, as shown in the radial profiles plotted in Figure 2c.

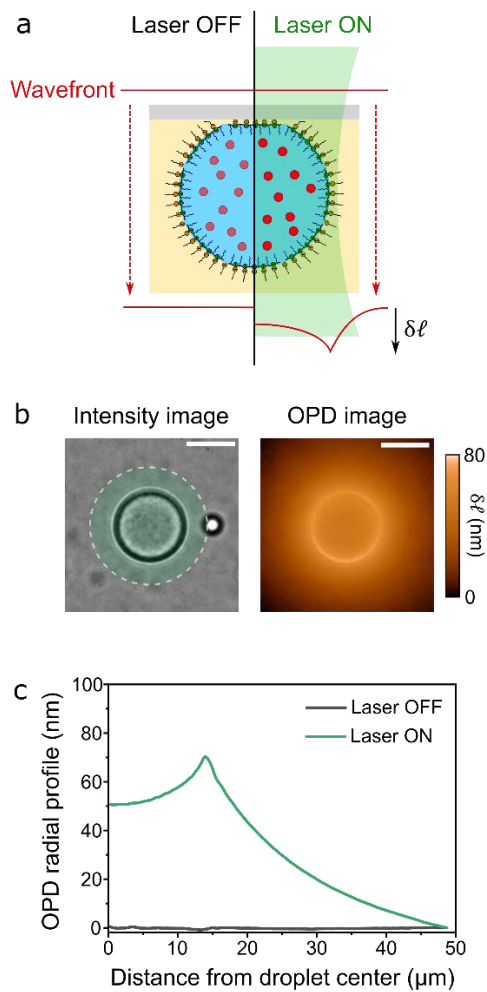

**Figure 2.** Thermoplasmonic phase signature, expressed in terms of optical path difference (OPD, $\delta\ell$), measured by cross-grating wavefront microscopy (CGM) on the 28-μm droplet under study in Figure 1. a) The droplet was imaged with a planar illumination at 625 nm coming from the top. Upon laser irradiation at 532 nm (coming from the bottom), the illumination wavefront was distorted due to temperature gradients induced by thermoplasmonic excitation of the AuNPs dispersed within the droplet. For the sake of readability, excess PLL-g-PNIPAM in the droplet core is not represented. b) Intensity and OPD images of the droplet provided by CGM, upon laser excitation at 12.7 μW/μm². To specifically detect the contribution of thermoplasmonic heating, an OPD image taken prior to laser irradiation was subtracted to subsequent images. The laser spot (50 μm in diameter) is schematically indicated in green in the intensity image. Scale bars = 20 μm. c) Radial OPD profiles (azimuthal average) measured on the droplet prior and upon laser excitation. The OPD origin ($\delta\ell = 0$) has been set arbitrarily far from the droplet center (here at ∼ 50 μm).

Due to the artificial offset in OPD values and to the absence of OPD/temperature analytical relationship in this three-dimensional configuration, absolute temperature values could not be directly retrieved from the experimental OPD profiles. To circumvent this limitation, the experimental data were compared to numerical data that were generated by simulations of steady-state heat transfers in axisymmetric geometry (corresponding to the experimental configuration). Because the AuNPs were uniformly distributed within the core of the droplet, we considered this latter as a uniform heat source dissipating a heating power $Q$ (**Figure S3** in SI). The relevant range of $Q$ values that should be used in the simulations (typically a fraction of mW) was determined from a rough estimation of the power dissipated in the experimental conditions (**Figure S4** in SI).

**Figure 3**a shows a side-view cross-section map of temperature rise generated by a 28-μm droplet in its environment (here for $Q = 0.20$ W). Knowing the $(\mathrm{d}n/\mathrm{d}T)$ coefficients of water and oil (Ref [7] and **Figure S5** in SI, respectively), the local temperature rise was converted into local variation of refractive index (Figure 3b). Eventually, the OPD radial profile was calculated by numerically integrating the variation of refractive index (taken positive) over the sample thickness (Figure 3c). In agreement with the experimental observations, the numerical OPD profiles exhibit a maximum located at the boundary of the droplet, resulting from a combination of (i) temperature gradients directed towards the droplet core and (ii) $|\mathrm{d}n/\mathrm{d}T|$ gradients directed towards the oil phase ($|\mathrm{d}n/\mathrm{d}T|_{\mathrm{water}} < |\mathrm{d}n/\mathrm{d}T|_{\mathrm{oil}}$). In addition, the numerical profiles capture the long-range decrease to zero of the OPD in the oil phase, that was not accessible in the experiments.

Performing these numerical simulations for different heating powers $Q$ eventually gave a direct correlation between heating power, temperature, and OPD values (e.g., the correlation between the heating power and the maximum/mean temperature rise in the droplet core is shown in Figure 3d).

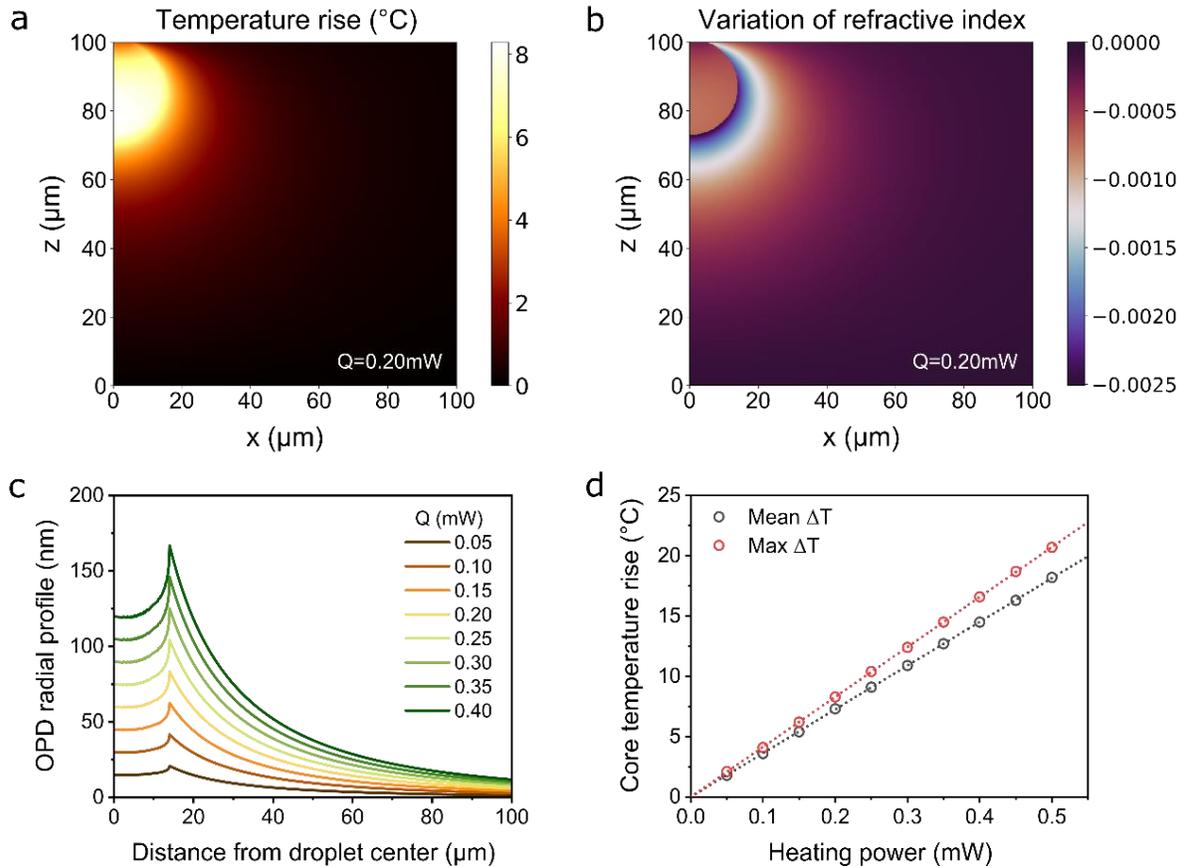

**Figure 3.** OPD radial profiles predicted from numerical solutions of the steady-state heat equation in axisymmetric geometry around a 28-µm heating droplet. a) Side-view cross-section map of the temperature rise $\Delta T$ induced by a heating power $Q$ = 0.20 mW uniformly distributed within the droplet. The simulated domain has been limited to the droplet and the surrounding oil phase on a length scale of 100 µm (corresponding to the internal capillary thickness). b) Side-view cross-section map of the variation of refractive index. c) OPD radial profiles calculated by numerical integration of the variation of refractive index along the $z$ direction for different values of heating power used in the simulation. d) Correlation between heating power and maximum/mean temperature rise reached in the droplet core (the mean temperature rise was obtained by an average on the core volume).

By comparison between experimental and numerical OPD profiles, one was able to infer the heating power that was dissipated in the experimental conditions, and then to deduce the temperature which was effectively reached in the droplet. This simulation-based inference protocol is shown in **Figure 4** for the droplet under study in Figure 1 and 2, using the numerical results shown in Figure 3. Experimental profiles were recorded for different values of laser irradiance, and were shifted to equal zero at a distance of 50 µm from the droplet center (Figure 4a, solid lines). Numerical OPD profiles were simulated for different heating powers (here regularly spaced between 0.05 and 0.40 mW), and shifted to cross zero at 50 µm from the droplet center, in order to be compared with the experimental OPD profiles (Figure 4a, dashed lines). The OPD values on the numerical profiles were used as an intrinsic OPD/heating power calibration. This is illustrated in Figure 4b by taking somewhat arbitrarily the value at 25 µm from the droplet center (i.e. in the oil phase), and projecting the experimental OPD values (measured at the same position) on this calibration curve to infer the heating power dissipated in the experiments (Figure 4b and Figure 4c). From the correlation between heating power and temperature provided by simulations (Figure 3d), the maximum temperature rise reached in the droplet could then be deduced for each irradiance condition (Figure 4c, right axis). For the droplet under study, an increase of 15 °C was thus determined at the maximum irradiance tested (20 µW/µm²).

When fed with the appropriate inferred values of heating power, simulations provided numerical profiles that nicely fitted the experimental OPD decay observed in the oil phase (Figure 4d). However, in the region corresponding to the droplet core the experimental OPD values were systematically exceeding the simulated values, especially at high irradiances. This discrepancy likely comes from the fact that the droplet was assimilated to pure water in the simulations, neglecting the presence of AuNPs and PLL-g-PNIPAM that changes the $(\mathrm{d}n/\mathrm{d}T)$ coefficient in the droplet. To match with experimental results, the effective value of the $(\mathrm{d}n/\mathrm{d}T)$ coefficient of the droplet should be more negative than that of water. We analyzed in SI (section 6) the variations with varying heating powers of the difference between numerical and experimental values of OPD. This analysis indicates that below the LCST of PLL-g-PNIPAM, the difference is constant and corresponds to a shift of about 30 % as compared to pure water (**Figure S6** in SI). The difference becomes more pronounced for temperatures above the LCST with an effective $(\mathrm{d}n/\mathrm{d}T)$ coefficient that is 2.4 times higher than that of pure water (Figure S6). Finally, the sensitivity of the thermal coefficient of aqueous solutions to composition, and possibly to temperature, suggests that it is more robust to rely on measurements and fits of temperature in the oil phase.

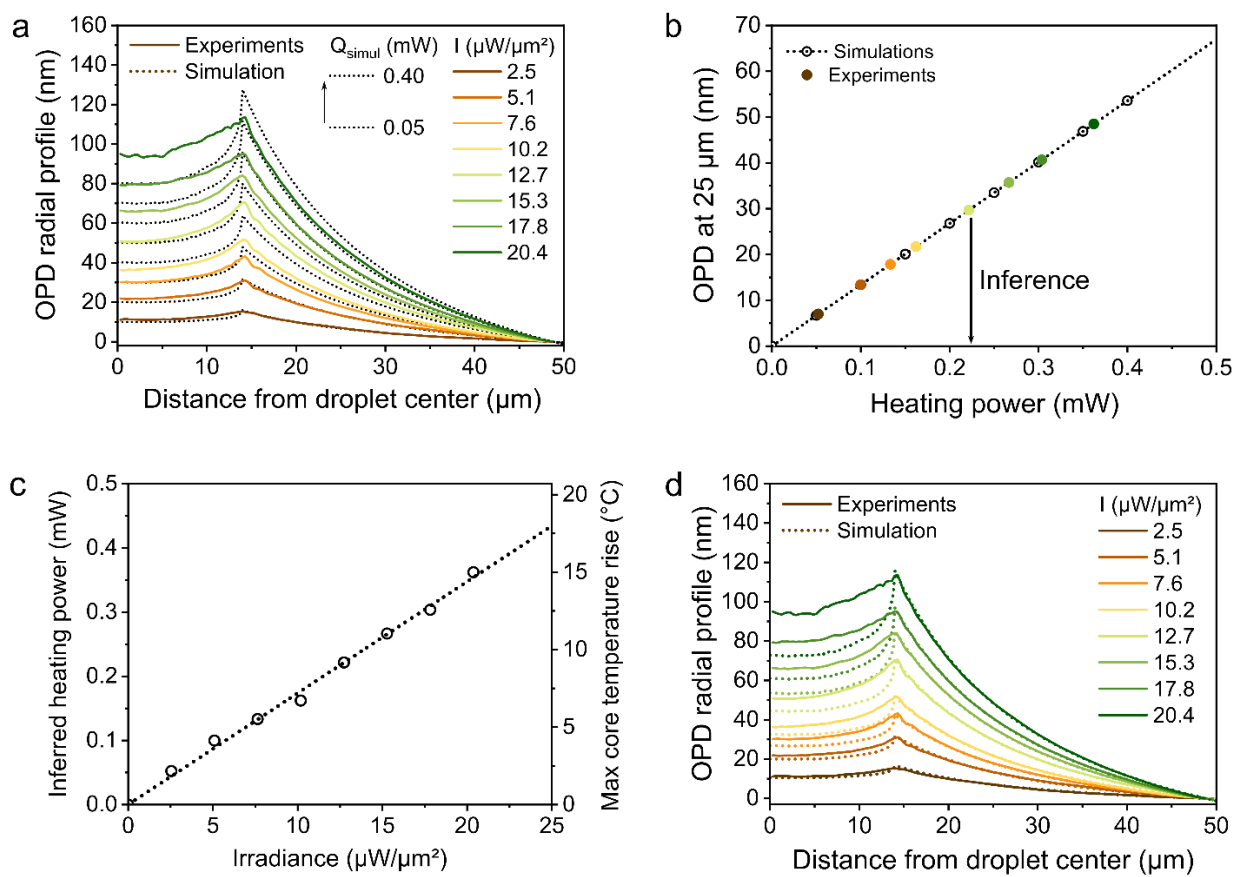

**Figure 4.** Simulation-based inference of the temperature rise in the droplet core from comparison between experimental and numerical OPD profiles. a) Comparison between experimental profiles (solid lines) and numerical profiles (dashed lines) generated using test-values of heating power (between 0.05 and 0.40 mW, with a fixed increment of 0.05 mW from bottom to top). An offset was applied to the numerical profiles to match with the same OPD origin than for the experimental profiles (here set at ∼ 50 µm). b) Numerical OPD values measured at 25 µm (black open circles) were used to plot an OPD/heating power calibration curve (dashed line). Projecting the experimental OPD values on this curve (closed colored circles) enabled to infer the effective values of heating power dissipated in the experiments. c) Evolution of the heating power inferred from the comparison between experimental and simulated profiles as a function of the laser irradiance. The corresponding maximum temperature rise reached in the droplet core (deduced from simulations, see Figure 3c) are shown on the right axis. d) Superimposition of the experimental OPD profiles (solid lines) and the numerical profiles generated with the inferred values of heating power (dashed lines).

The inference-based thermometry approach described in Figure 4 was performed on droplets of different sizes to quantify the absolute maximal temperature reached in the droplet core upon laser excitation at different irradiances (**Figure 5**). Starting from room temperature (here 22 °C), the temperature was risen up to 29 °C, 31.5 °C, and 37 °C in droplets of diameter 18 μm, 22 μm, and 28 μm respectively. This evolution is consistent with the fact that the number of AuNPs in the droplet core, and consequently the heating power, increases with the droplet size (the concentration of AuNPs being fixed). Interestingly, we noticed that the LCST of PLL-g-PNIPAM (33 °C) was exceeded in the 28-μm droplet for irradiances greater than 15 μW/μm². This result was in agreement with the qualitative observation of the collapse of PLL-g-PNIPAM as described in Figure 1, validating the quantitative thermometry approach presented throughout this article. In addition, since this approach is based on the measurement of the OPD in the oil phase surrounding the droplet, it does not depend on the composition (and potential heterogeneities) in the droplet, and does not require any internal probe to be carried out. This process could thus be applied to any three-dimensional thermoplasmonic system whose geometry can be numerically simulated.

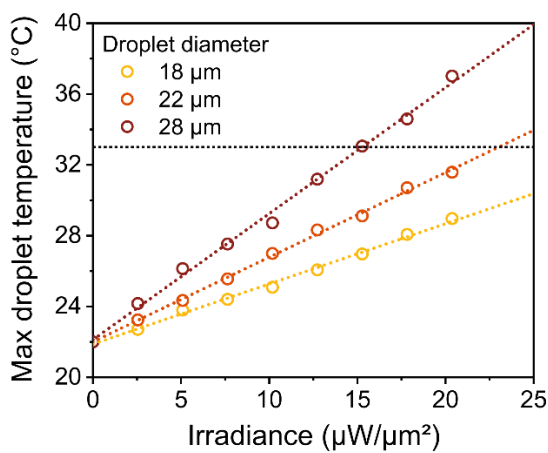

**Figure 5.** Absolute maximal temperature reached in the droplet core as deduced from simulation-based inference for droplets of different diameters (room temperature was set at 22 °C). The horizontal black dashed line corresponds to the LCST of PLL-g-PNIPAM (33 °C).

## Conclusion

In summary, this article introduces a label-free thermometry approach that enables the measurement of *in situ* temperature fields generated at the microscale by three-dimensional thermoplasmonic sources upon laser excitation. This technique is based on the measurement of heat-induced variations of optical path difference (OPD) in the sample by cross-grating wavefront microscopy. This approach was tested on a model system consisting in emulsion droplets loaded with gold nanoparticles, for which the thermoplasmonic-induced OPD signature cannot be straightforwardly converted into temperature values using analytical tools. Numerical simulations of steady-state heat transfers were used to predict the relationship between optical path difference and temperature fields in the geometry of the system. A simulation-based inference protocol comparing experimental and numerical data eventually yields the temperature effectively reached in the experimental conditions. Typically, this approach enabled to measure temperature rises of 15 °C in ~30-μm droplets under laser excitation up to 20 μW/μm². The droplets were also loaded with an extrinsic thermosensitive probe, namely a water-soluble thermoresponsive polymer exhibiting a collapse transition towards microaggregates at 33 °C. Experimentally, this transition was observed when the temperature measured by our quantitative approach precisely exceeded 33 °C, confirming its potential for accurate *in situ* thermometry. Being non-invasive, label-free, robust, and versatile, we believe that this approach based on wavefront microscopy could be applied to many three-dimensional thermoplasmonic sources, including soft and biological systems loaded with AuNPs such as hydrogels,[47] vesicles,[48] polymer membranes,[49] or even living cells,[50] for which measuring three-dimensional temperature fields can be crucial in terms of applications.

## ASSOCIATED CONTENT

**Supporting Information**. Detailed experimental procedures (including the synthesis of PLL-g-PNIPAM, the formulation of the emulsions, and details on the wavefront microscopy set-up) and supplementary Figures S1-S6 (including the determination of the LCST of PLL-g-PNIPAM, emulsion micrographs, details on the simulation procedure, an estimation of experimental values of heating power, the measurement of $(dn/dT)$ of the oil phase, and the experimental determination of the effective $(dn/dT)$ in the droplet core).

## AUTHOR INFORMATION

Corresponding Authors

Lucas Sixdenier – PASTEUR, Département de chimie, École normale supérieure, PSL University, Sorbonne Université, CNRS, 75005 Paris, France. *Present address*: Physique et Mécanique des Milieux Hétérogènes, PMMH, ESPCI Paris, PSL University, CNRS, Sorbonne

Université, Université de Paris, 10, rue Vauquelin, 75005 Paris, France. ORCID: orcid.org/0000-0002-7085-0970. Email: lucas.sixdenier@espci.psl.eu

Emmanuelle Marie – PASTEUR, Département de chimie, École normale supérieure, PSL University, Sorbonne Université, CNRS, 75005 Paris, France. Email: emmanuelle.marie@ens.psl.eu

Author Contributions

**Lucas Sixdenier**: Conceptualization, Methodology, Investigation, Formal analysis, Validation, Visualization, Writing. **Guillaume Baffou**: Conceptualization, Methodology, Formal analysis, Writing. **Christophe Tribet**: Conceptualization, Writing, Supervision, Funding acquisition. **Emmanuelle Marie**: Conceptualization, Writing, Supervision, Funding acquisition.

Funding Sources

This work was supported by the following institutions and programs: the DYNAMO Laboratory of Excellence (ANR-11-LABEX-0011-01); the Agence Nationale pour la Recherche (ANR-20-CE06-0007 Fliposomes and ANR-17-CE09-0019 CASCADE); and Sorbonne University (Ecole Doctorale 388 scholarship).

Notes

## ACKNOWLEDGMENT

We thank Ludovic Jullien for seminal discussions, including on thermo-fluorimetry; Sadman Shakib for having measured the refractive index of the oil phase; and Victor Chardès for his valuable insights into numerical methods.

## ABBREVIATIONS

AuNPs, gold nanoparticles; CGM, cross-grating wavefront microscopy; LCST, lower critical solution temperature; OPD, optical path difference; PLL, poly(L-lysine); PNIPAM, poly(N-isopropylacrylamide).

## REFERENCES

(1)     Brongersma, M. L.; Halas, N. J.; Nordlander, P. Plasmon-Induced Hot Carrier Science and Technology. *Nat. Nanotechnol.* **2015**, *10* (1), 25–34. https://doi.org/10.1038/nnano.2014.311.

(2)     Amendola, V.; Pilot, R.; Frasconi, M.; Maragò, O. M.; Iatì, M. A. Surface Plasmon Resonance in Gold Nanoparticles: A Review. *J. Phys. Condens. Matter* **2017**, *29* (20). https://doi.org/10.1088/1361-648X/aa60f3.

(3)     Liz-Marzan, L. M. Tailoring Surface Plasmons through the Morphology and Assembly of Metal Nanoparticles. *Langmuir* **2006**, *22* (1), 32–41.

(4)     Baffou, G.; Quidant, R.; Girard, C. Heat Generation in Plasmonic Nanostructures: Influence of Morphology. *Appl. Phys. Lett.* **2009**, *94* (15). https://doi.org/10.1063/1.3116645.

(5)     Zhao, P.; Li, N.; Astruc, D. State of the Art in Gold Nanoparticle Synthesis. *Coord. Chem. Rev.* **2013**, *257* (3–4), 638–665. https://doi.org/10.1016/j.ccr.2012.09.002.

(6)     Liz-Marzán, L. M. *Colloidal Synthesis of Plasmonic Nanometals*, 1st Editio.; Jenny Stanford Publishing, 2020. https://doi.org/https://doi.org/10.1201/9780429295188.

(7)     Baffou, G. *Thermoplasmonics: Heating Metal Nanoparticles Using Light*; Cambridge University Press, 2017. https://doi.org/10.1017/9781108289801.

(8)     Baffou, G.; Cichos, F.; Quidant, R. Applications and Challenges of Thermoplasmonics. *Nat. Mater.* **2020**, *19* (9), 946–958. https://doi.org/10.1038/s41563-020-0740-6.

(9)     Cai, K.; Zhang, W.; Zhang, J.; Li, H.; Han, H.; Zhai, T. Design of Gold Hollow Nanorods with Controllable Aspect Ratio for Multimodal Imaging and Combined Chemo-Photothermal Therapy in the Second Near-Infrared Window. *ACS Appl. Mater. Interfaces* **2018**, *10* (43), 36703–36710. https://doi.org/10.1021/acsami.8b12758.

(10)    Wu, Y.; Ali, M. R. K.; Dong, B.; Han, T.; Chen, K.; Chen, J.; Tang, Y.; Fang, N.; Wang, F.; El-Sayed, M. A. Gold Nanorod Photothermal Therapy Alters Cell Junctions and Actin Network in Inhibiting Cancer Cell Collective Migration. *ACS Nano* **2018**, *12* (9), 9279–9290. https://doi.org/10.1021/acsnano.8b04128.

(11)    Song, J.; Fang, Z.; Wang, C.; Zhou, J.; Duan, B.; Pu, L.; Duan, H. Photolabile Plasmonic Vesicles Assembled from Amphiphilic Gold Nanoparticles for Remote-Controlled Traceable Drug Delivery. *Nanoscale* **2013**, *5* (13), 5816–5824. https://doi.org/10.1039/c3nr01350b.

(12)    Shao, J.; Xuan, M.; Si, T.; Dai, L.; He, Q. Biointerfacing Polymeric Microcapsules for in

Vivo Near-Infrared Light-Triggered Drug Release. *Nanoscale* **2015**, *7* (45), 19092–19098. https://doi.org/10.1039/c5nr06350g.

(13) Zhu, M.; Baffou, G.; Meyerbröker, N.; Polleux, J. Micropatterning Thermoplasmonic Gold Nanoarrays to Manipulate Cell Adhesion. *ACS Nano* **2012**, *6* (8), 7227–7233. https://doi.org/10.1021/nn302329c.

(14) Bahadori, A.; Oddershede, L. B.; Bendix, P. M. Hot-Nanoparticle-Mediated Fusion of Selected Cells. *Nano Res.* **2017**, *10* (6), 2034–2045. https://doi.org/10.1007/s12274-016-1392-3.

(15) Liu, Z.; Liu, Y.; Chang, Y.; Seyf, H. R.; Henry, A.; Mattheyses, A. L.; Yehl, K.; Zhang, Y.; Huang, Z.; Salaita, K. Nanoscale Optomechanical Actuators for Controlling Mechanotransduction in Living Cells. *Nat. Methods* **2016**, *13* (2), 143–146. https://doi.org/10.1038/nmeth.3689.

(16) Baffou, G.; Quidant, R. Nanoplasmonics for Chemistry. *Chem. Soc. Rev.* **2014**, *43* (11), 3898–3907. https://doi.org/10.1039/c3cs60364d.

(17) Robert, H. M. L.; Kundrat, F.; Bermúdez-Ureña, E.; Rigneault, H.; Monneret, S.; Quidant, R.; Polleux, J.; Baffou, G. Light-Assisted Solvothermal Chemistry Using Plasmonic Nanoparticles. *ACS Omega* **2016**, *1* (1), 2–8. https://doi.org/10.1021/acsomega.6b00019.

(18) Cortés, E.; Grzeschik, R.; Maier, S. A.; Schlücker, S. Experimental Characterization Techniques for Plasmon-Assisted Chemistry. *Nat. Rev. Chem.* **2022**, *6* (4), 259–274. https://doi.org/10.1038/s41570-022-00368-8.

(19) Powell, A. W.; Stavrinadis, A.; De Miguel, I.; Konstantatos, G.; Quidant, R. White and

Brightly Colored 3D Printing Based on Resonant Photothermal Sensitizers. *Nano Lett.* **2018**, *18* (11), 6660–6664. https://doi.org/10.1021/acs.nanolett.8b01164.

(20)   Quintanilla, M.; Liz-Marzán, L. M. Guiding Rules for Selecting a Nanothermometer. *Nano Today* **2018**, *19*, 126–145. https://doi.org/10.1016/j.nantod.2018.02.012.

(21)   Chiu, M. J.; Chu, L. K. Quantifying the Photothermal Efficiency of Gold Nanoparticles Using Tryptophan as an in Situ Fluorescent Thermometer. *Phys. Chem. Chem. Phys.* **2015**, *17* (26), 17090–17100. https://doi.org/10.1039/c5cp02620b.

(22)   Bendix, P. M.; Reihani, S. N. S.; Oddershede, L. B. Direct Measurements of Heating by Electromagnetically Trapped Gold Nanoparticles on Supported Lipid Bilayers. *ACS Nano* **2010**, *4* (4), 2256–2262. https://doi.org/10.1021/nn901751w.

(23)   Ebrahimi, S.; Akhlaghi, Y.; Kompany-Zareh, M.; Rinnan, Å. Nucleic Acid Based Fluorescent Nanothermometers. *ACS Nano* **2014**, *8* (10), 10372–10382. https://doi.org/10.1021/nn5036944.

(24)   Coppens, Z. J.; Li, W.; Walker, D. G.; Valentine, J. G. Probing and Controlling Photothermal Heat Generation in Plasmonic Nanostructures. *Nano Lett.* **2013**, *13* (3), 1023–1028. https://doi.org/10.1021/nl304208s.

(25)   Freddi, S.; Sironi, L.; D'Antuono, R.; Morone, D.; Donà, A.; Cabrini, E.; D'Alfonso, L.; Collini, M.; Pallavicini, P.; Baldi, G.; Maggioni, D.; Chirico, G. A Molecular Thermometer for Nanoparticles for Optical Hyperthermia. *Nano Lett.* **2013**, *13* (5), 2004–2010. https://doi.org/10.1021/nl400129v.

(26)   Aigouy, L.; Tessier, G.; Mortier, M.; Charlot, B. Scanning Thermal Imaging of

Microelectronic Circuits with a Fluorescent Nanoprobe. *Appl. Phys. Lett.* **2005**, *87* (18), 1–3. https://doi.org/10.1063/1.2123384.

(27)    Baffou, G.; Kreuzer, M. P.; Kulzer, F.; Quidant, R. Temperature Mapping near Plasmonic Nanostructures Using Fluorescence Polarization Anisotropy. *Opt. Express* **2009**, *17* (5), 3291. https://doi.org/10.1364/oe.17.003291.

(28)    Donner, J. S.; Thompson, S. A.; Kreuzer, M. P.; Baffou, G.; Quidant, R. Mapping Intracellular Temperature Using Green Fluorescent Protein. *Nano Lett.* **2012**, *12* (4), 2107–2111. https://doi.org/10.1021/nl300389y.

(29)    Bradac, C.; Lim, S. F.; Chang, H. C.; Aharonovich, I. Optical Nanoscale Thermometry: From Fundamental Mechanisms to Emerging Practical Applications. *Adv. Opt. Mater.* **2020**, *8* (15), 1–29. https://doi.org/10.1002/adom.202000183.

(30)    Blum, O.; Shaked, N. T. Prediction of Photothermal Phase Signatures from Arbitrary Plasmonic Nanoparticles and Experimental Verification. *Light Sci. Appl.* **2015**, *4* (8). https://doi.org/10.1038/lsa.2015.95.

(31)    Baffou, G. Quantitative Phase Microscopy Using Quadriwave Lateral Shearing Interferometry (QLSI): Principle, Terminology, Algorithm and Grating Shadow Description. *J. Phys. D. Appl. Phys.* **2021**, *54* (29). https://doi.org/10.1088/1361-6463/abfbf9.

(32)    Marthy, B.; Baffou, G. Cross-Grating Phase Microscopy (CGM): In-Silico Experiments, Noise and Accuracy. *Opt. Commun.* **2022**, *521*, 128577. https://doi.org/10.1016/j.optcom.2022.128577.

(33)    Baffou, G.; Bon, P.; Savatier, J.; Polleux, J.; Zhu, M.; Merlin, M.; Rigneault, H.; Monneret, S. Thermal Imaging of Nanostructures by Quantitative Optical Phase Analysis. *ACS Nano* **2012**, *6* (3), 2452–2458. https://doi.org/10.1021/nn2047586.

(34)    Baffou, G.; Berto, P.; Bermúdez Ureña, E.; Quidant, R.; Monneret, S.; Polleux, J.; Rigneault, H. Photoinduced Heating of Nanoparticle Arrays. *ACS Nano* **2013**, *7* (8), 6478–6488. https://doi.org/10.1021/nn401924n.

(35)    Bon, P.; Belaid, N.; Lagrange, D.; Bergaud, C.; Rigneault, H.; Monneret, S.; Baffou, G. Three-Dimensional Temperature Imaging around a Gold Microwire. *Appl. Phys. Lett.* **2013**, *102* (24), 1–5. https://doi.org/10.1063/1.4811557.

(36)    Berto, P.; Philippet, L.; Osmond, J.; Liu, C. F.; Afridi, A.; Montagut Marques, M.; Molero Agudo, B.; Tessier, G.; Quidant, R. Tunable and Free-Form Planar Optics. *Nat. Photonics* **2019**, *13* (9), 649–656. https://doi.org/10.1038/s41566-019-0486-3.

(37)    Cheng, G.; Lin, K. T.; Ye, Y.; Jiang, H.; Ngai, T.; Ho, Y. P. Photo-Responsive Fluorosurfactant Enabled by Plasmonic Nanoparticles for Light-Driven Droplet Manipulation. *ACS Appl. Mater. Interfaces* **2021**, *13* (18), 21914–21923. https://doi.org/10.1021/acsami.0c22900.

(38)    Phan-Quang, G. C.; Lee, H. K.; Phang, I. Y.; Ling, X. Y. Plasmonic Colloidosomes as Three-Dimensional SERS Platforms with Enhanced Surface Area for Multiphase Sub-Microliter Toxin Sensing. *Angew. Chemie* **2015**, *127* (33), 9827–9831. https://doi.org/10.1002/ange.201504027.

(39)    Wei, S. C.; Hsu, M. N.; Chen, C. H. Plasmonic Droplet Screen for Single-Cell Secretion

Analysis. *Biosens. Bioelectron.* **2019**, *144* (August), 111639. https://doi.org/10.1016/j.bios.2019.111639.

(40) Huang, Y.; Huang, P.; Lin, J. Plasmonic Gold Nanovesicles for Biomedical Applications. *Small Methods* **2019**, *3* (3), 1–17. https://doi.org/10.1002/smtd.201800394.

(41) Sixdenier, L.; Augé, A.; Zhao, Y.; Marie, E.; Tribet, C. UCST-Type Polymer Capsules Formed by Interfacial Complexation. *ACS Macro Lett.* **2022**, No. ii, 651–656. https://doi.org/10.1021/acsmacrolett.2c00021.

(42) Dejournette, C. J.; Kim, J.; Medlen, H.; Li, X.; Vincent, L. J.; Easley, C. J. Creating Biocompatible Oil-Water Interfaces without Synthesis: Direct Interactions between Primary Amines and Carboxylated Perfluorocarbon Surfactants. *Anal. Chem.* **2013**, *85* (21), 10556–10564. https://doi.org/10.1021/ac4026048.

(43) Sixdenier, L.; Tribet, C.; Marie, E. Emulsion-Templated Poly(N-Isopropylacrylamide) Shells Formed by Thermo-Enhanced Interfacial Complexation. *Adv. Funct. Mater.* **2021**, *2105490* (in press). https://doi.org/10.1002/adfm.202105490.

(44) Aibara, I.; Chikazawa, J. I.; Uwada, T.; Hashimoto, S. Localized Phase Separation of Thermoresponsive Polymers Induced by Plasmonic Heating. *J. Phys. Chem. C* **2017**, *121* (40), 22496–22507. https://doi.org/10.1021/acs.jpcc.7b07187.

(45) Iwai, K.; Matsumura, Y.; Uchiyama, S.; De Silva, A. P. Development of Fluorescent Microgel Thermometers Based on Thermo-Responsive Polymers and Their Modulation of Sensitivity Range. *J. Mater. Chem.* **2005**, *15* (27–28), 2796–2800. https://doi.org/10.1039/b502277k.

(46)     Ding, Z.; Wang, C.; Feng, G.; Zhang, X. Thermo-Responsive Fluorescent Polymers with Diverse LCSTs for Ratiometric Temperature Sensing through FRET. *Polymers (Basel).* **2018**, *10* (3). https://doi.org/10.3390/polym10030283.

(47)     Zhao, J.; Su, H.; Vansuch, G. E.; Liu, Z.; Salaita, K.; Dyer, R. B. Localized Nanoscale Heating Leads to Ultrafast Hydrogel Volume-Phase Transition. *ACS Nano* **2019**, *13* (1), 515–525. https://doi.org/10.1021/acsnano.8b07150.

(48)     Amstad, E.; Kim, S. H.; Weitz, D. A. Photo- and Thermoresponsive Polymersomes for Triggered Release. *Angew. Chemie - Int. Ed.* **2012**, *51* (50), 12499–12503. https://doi.org/10.1002/anie.201206531.

(49)     Hu, L.; Gao, S.; Ding, X.; Wang, D.; Jiang, J.; Jin, J.; Jiang, L. Photothermal-Responsive Single-Walled Carbon Nanotube-Based Ultrathin Membranes for on/off Switchable Separation of Oil-in-Water Nanoemulsions. *ACS Nano* **2015**, *9* (5), 4835–4842. https://doi.org/10.1021/nn5062854.

(50)     Li, M.; Lohmüller, T.; Feldmann, J. Optical Injection of Gold Nanoparticles into Living Cells. *Nano Lett.* **2015**, *15* (1), 770–775. https://doi.org/10.1021/nl504497m.

TABLE OF CONTENTS

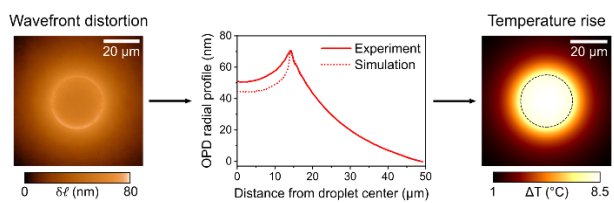

Supporting information

# Quantitative Microscale Thermometry in Droplets Loaded with Gold Nanoparticles

*Lucas Sixdenier,[a,]\* Guillaume Baffou,[b] Christophe Tribet,[a] Emmanuelle Marie [a,]\**

[a] PASTEUR, Département de chimie, École normale supérieure, PSL University, Sorbonne Université, CNRS, 75005 Paris, France

[b] Institut Fresnel, CNRS, Aix Marseille University, Centrale Marseille, Marseille, France

# Content

# I.   Materials and methods

## 1.   Materials

Poly(L-lysine) hydrobromide (PLL, 15-30 kDa), (N-hydroxysuccinimidyl ester)-terminated poly(N-isopropylacrylamide) (PNIPAM-NHS, 2 kDa), and Fluorinert FC-70 oil were purchased from Sigma-Aldrich. Krytox 157 FS(L) ($M_W$ = 2.5 kDa) was purchased from Samaro. Econix$^{TM}$ gold nanoparticles coated with 40 kDa poly(vinyl pyrrolidone) (50 nm diameter, 5 mg/mL in water) were purchased from nanoComposix. Unless otherwise noticed, all chemicals were used without further purification.

## 2.   Synthesis of poly(L-lysine)-g-poly(N-isopropylacrylamide) (PLL-g-PNIPAM)

PLL-g-PNIPAM was prepared according to a protocol described in Ref [1]. Briefly, PLL hydrobromide ($M_W$ = 15-30 kg/mol, 15 mg) was dissolved in sodium tetraborate buffer (5 mL, pH = 8.5). PNIPAM-NHS ($M_W$ = 2 kg/mol, 48 mg) was then added and the solution was stirred at 4 °C until complete dissolution of the polymer. Then, the solution was stirred for an additional 4h at room temperature. The resulting solution was dialyzed against deionized water for 3 days in a Thermo Scientific Slide-A-Lyzer cassette (3-12 mL, MW cutoff = 3.5 kDa) and finally freeze-dried for 2 days in a Labconco Freezone Plus 2.5 apparatus. The effective grafting ratio in the final PLL-g-PNIPAM was around 0.20.

# 3.    Measurement of the LCST of PLL-g-PNIPAM by turbidimetry

The turbidimetry curve was deduced from the measurement of the transmittance of a polymer solution in the visible range. Transmittance spectra were recorded on a single cell Thermo Scientific Evolution Array UV-Vis spectrophotometer equipped with a Peltier temperature-controlled cell holder (+/− 0.1 °C). The solution of PLL-g-PNIPAM (at 20 g/L) was injected in a 150 µL quartz micro-cuvette (length 1 cm) and submitted to an increase of temperature from 20 °C to 60 °C by steps of 2 °C. A spectrum was recorded at every step temperature after 3 min thermalization, and the mean transmittance between 650 and 750 nm was used to plot the turbidimetry curve.

# 4.    Formulation of the emulsions

5 vol% water-in-oil emulsions were prepared as follows: 5 µL of aqueous phase (containing PLL-g-PNIPAM at 20 g/L, and AuNPs at 0.5 g/L in 7.7 mM phosphate buffer, pH = 7.3) and 95 µL of FC-70 oil phase (containing Krytox at 0.5 g/L) were mixed in a 0.5 mL Eppendorf tube and manually emulsified until the mixture became turbid. For microscopy experiments, a few µL of emulsion were injected in a hollow rectangle capillary (H × W = 0.1 × 1 mm, CMScientific) previously coated with PLL-g-PEG (incubated at 1 g/L in 7.7 mM phosphate buffer) to prevent the breakage of water droplets at the contact with bare glass side. After injection of the emulsion, the capillary was sealed with Vitrex putty.

## 5.   Cross-grating wavefront microscopy

Cross-grating wavefront microscopy (CGM) was performed on a home-made optical set-up (see **Figure i**), including: a 700 mW LED illumination operating at 625 nm (M625L3, Thorlabs), a 60x objective (LUC Plan FLN, NA 0.7, Olympus), and a CGM wavefront sensor (SID4-sC8, Phasics). The plasmonic excitation of the sample was achieved with a diode-pumped solid-state laser operating at 532 nm (MGL-FN-532 DPSS, CNI Optoelectronics Tech). The diameter of the laser spot in the focal plane of the objective was set to 50 µm thanks to a beam-expander set-up, thus ensuring a uniform plasmonic excitation in droplets with a diameter up to 50 µm. The laser power delivered to the sample was set between 0 and ~ 40 mW, corresponding to an irradiance (i.e. a power per unit area) between 0 and ~ 20 µW/µm². The droplets were imaged at a frame rate of 1 Hz, prior and after turning the laser on. An image recorded just before turning the laser on was subtracted to the images recorded upon laser irradiation to isolate and measure the optical path difference due to thermoplasmonic effects.

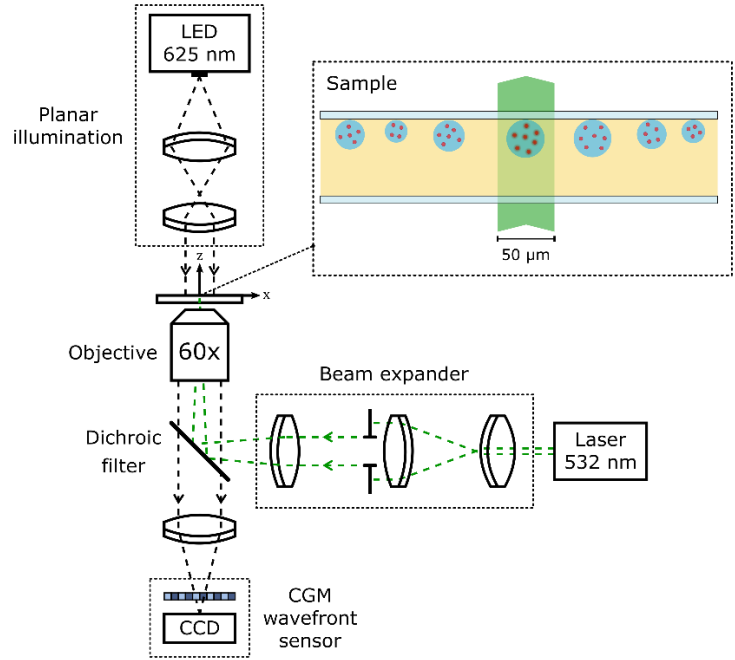

**Figure i**. Experimental cross-grating wavefront microscopy set-up. The sample is a water-in-oil emulsion injected in a glass capillary. The water droplets (blue disks) were loaded with 20 g/L of PLL-g-PNIPAM (not represented) and 0.5 g/L AuNPs (red dots). The sample was illuminated by the top with a 625-nm diode, and excited by the bottom with a 532-nm laser (in green).

# II. Supplementary figures

## 1. Turbidimetry curve of PLL-g-PNIPAM at 20 g/L

The transmittance of a solution of PLL-g-PNIPAM at 20 g/L in phosphate buffer (7.7 mM, pH = 7.3) was measured upon heating from 20 °C to 50 °C (see section I.3 above). The sharp decrease of the transmittance between 32°C and 35 °C evidences the LCST behavior of the PLL-g-PNIPAM (i.e. the aggregative transition of the polymer chains). The cloud point of the solution (i.e. the temperature at which the solution becomes turbid) was measured at the intersection between the tangent to the inflexion point of the transmittance curve and the initial plateau at 100 % transmittance, and was equal to 33 °C. By misuse of language, this cloud point value was assimilated to the LCST of the polymer (which rigorously corresponds to the cloud point obtained at the minimum of the binodal curve, i.e. for a precise concentration in polymer).[2]

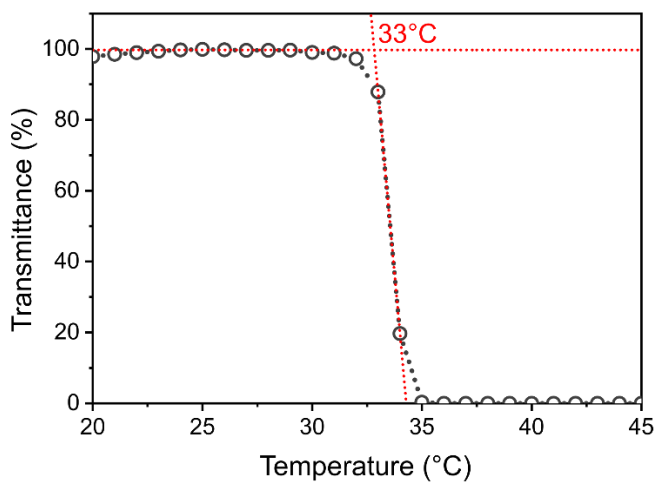

**Figure S1**. Turbidity heating curve of a PLL-g-PNIPAM solution at 20 g/L in 7.7 mM phosphate buffer.

## 2.    Top- and side view of an emulsion injected in a capillary

**Figure S2** shows top- and side-view micrographs of an emulsion prepared and injected in a rectangle capillary as described above in section I.4. The side-view micrographs show that droplets are slightly flattened (typically over 1 µm) at the contact point with the internal upper side of the capillary.

For CGM experiments, the emulsion was diluted to get quasi-isolated droplets, at least at the scale of the imaging field of view.

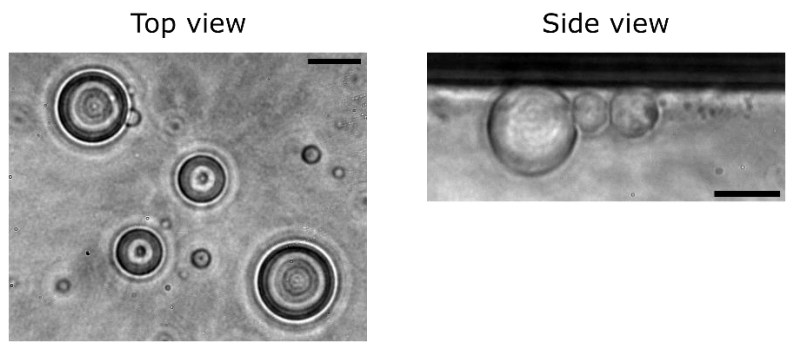

**Figure S2**. Side- and top-view micrographs of an emulsion injected in a capillary (scale bars = 30 µm).

## 3.    Details on the simulation procedure

Heat transfer simulations were performed with COMSOL Multiphysics®. The simulation domain was defined as an oil layer (length = 1000 µm, thickness = 100 µm) sandwiched between two glass spherical caps (radius of the cap base = 1000 µm, height = 500 µm). An isolated water droplet (radius $R$) was placed in the oil phase, and was flattened over a distance $e$ at the contact with the upper capillary side to match the experimental observation shown in Figure S2. **Figure S3** shows a side-view cross-section of the simulation domain in axisymmetric geometry (zoomed at the level of the droplet), including the mesh grid used to solve the heat equation. The droplet was defined as a uniform heat source dissipating a heating power $Q$. Knowing the relevant physical parameters of the different media (see **Table 1**), and setting the boundary condition for the temperature far from the droplet ($T = 275$ K at the surface delimiting the whole domain), the heat equation $-\kappa \Delta T = Q/V$ was numerically solved in steady state over the whole domain.

The simulation provided side-view cross-section temperature maps that could be converted into maps of variation of refractive index, whose integration along the height of the capillary eventually generated radial OPD profiles.

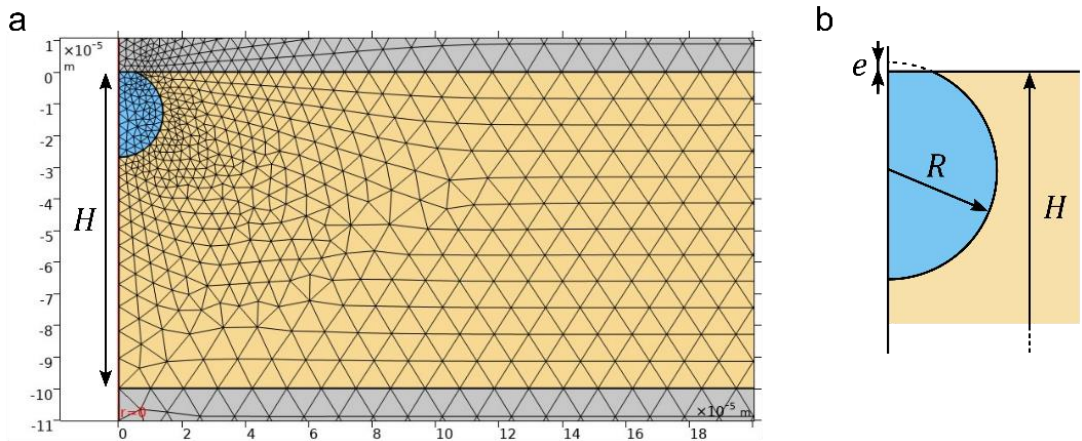

**Figure S3**. Representation of the simulation domain. a) Side-view representation of the simulation in axisymmetric geometry, and of the integration mesh used to solve the heat equation (grey = glass capillary borders, blue = water droplet, yellow = oil phase). The internal thickness of the capillary containing the emulsion is $H = 100$ μm. b) Zoom on the droplet to highlight the droplet radius $R$ and the flattening distance $e$ at the contact with the upper internal side of the capillary.

| | Thermal conductivity ($\mathrm{W.m^{-1}.K^{-1}}$) | Heat capacity at constant pressure ($\mathrm{kJ.kg^{-1}.K^{-1}}$) | Density ($\mathrm{kg.m^{-3}}$) |
|---|---|---|---|
| Glass | 1 | 840 | 2500 |
| FC-70 oil | 0.07 | 1.1 | 1940 |
| Water | 0.6 | 4.2 | 1000 |

**Table 1**. Physical parameters used in the simulation.

## 4.  Estimation of the heating power dissipated in the experiments.

To estimate the relevant range of heat power to be set in the simulation, the heat power experimentally generated in the droplets was derived from simple order of magnitude considerations regarding the formulation of the droplets and the conditions of irradiation. We assumed that the total heat power $Q$ which was dissipated in a droplet was the sum of the heat power dissipated by each individual AuNP loaded in the droplet. Hence, in a droplet of volume $V$ containing a concentration of AuNPs $C_{AuNPs}$ (0.5 g/L = 0.45 NP/$\mu m^3$ according to the datasheet) and uniformly irradiated by an irradiance $I$, $Q$ was estimated by the following formula: $Q = C_{AuNPs}V\sigma_{abs}I$, where $\sigma_{abs}$ is the absorption cross-section of a single AuNP, which is equal to $6 \times 10^{-3}$ nm$^2$ for a AuNP with a diameter of 50 nm.[3]

**Figure S4** shows that the heat power dissipated in droplets of different size during the irradiation experiments is of the order of magnitude of a fraction of mW. This is typically the values of heating power that were tested in the simulation ($\sim 0.01 - 0.6$ mW).

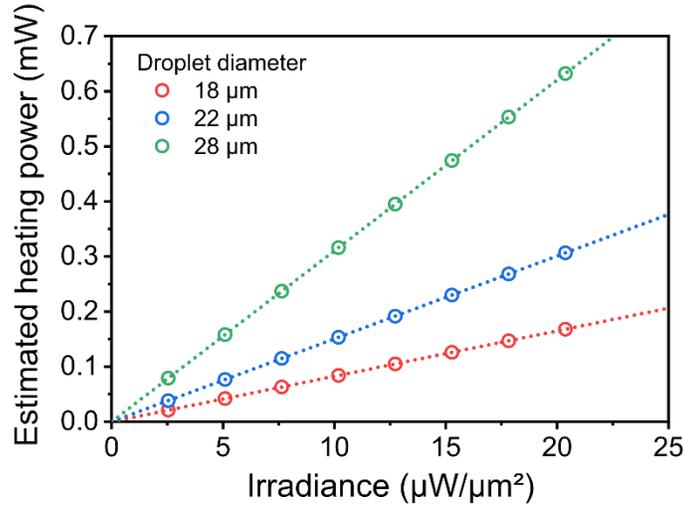

**Figure S4**. Estimation of the heat power dissipated in droplets of different sizes upon uniform irradiation at 532 nm as a function of the laser irradiance.

## 5.     Measurement of the $(\mathrm{d}n/\mathrm{d}T)$ coefficient of the oil phase

The $(\mathrm{d}n/\mathrm{d}T)$ coefficient of FC-70 oil was measured using cross-grating wavefront microscopy. Briefly, a 12 μm-deep crater etched in a glass slide was filled with oil. An OPD image of the crater was taken at different temperatures upon heating between room temperature (22 °C) and 105 °C (**Figure S5**a shows the OPD image at 90 °C). At each temperature, the refractive index of the oil $n_O$ was deduced from the measurement of the OPD $(\delta\ell)$,[4] using the following formula: $n_O = n_{air} + \delta\ell/h$, with $n_{air} = 1.00$ the refractive index of the air, and $h = 12$ μm the depth of the crater. The evolution of $n_O$ with the temperature is shown in Figure S5b. The slope of this curve gives $(\mathrm{d}n/\mathrm{d}T) = -3.16 \times 10^{-4}$ $\mathrm{K}^{-1}$.

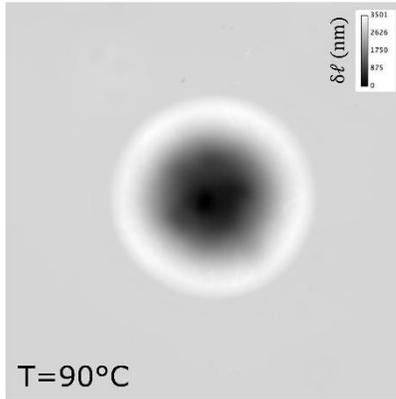

a  12-µm crater filled with FC-70

T=90°C

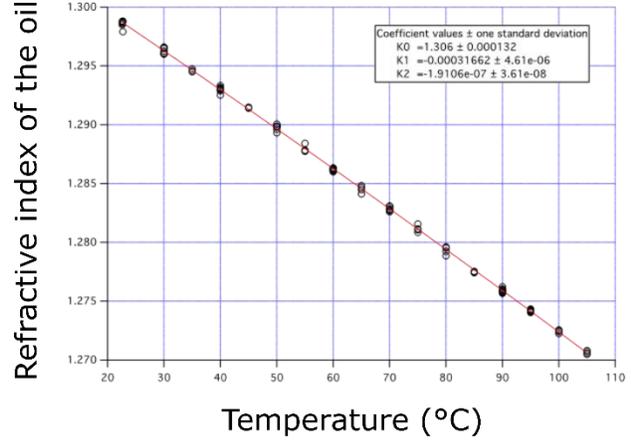

b

Refractive index of the oil

Temperature (°C)

**Figure S5**. Determination of the $(\mathrm{d}n/\mathrm{d}T)$ coefficient of FC-70 oil by CGM. a) OPD image of a 12-µm crater filled with FC-70 oil, heated at 90 °C. The refractive index of the oil was deduced from the measurement of the OPD generated by the crater. b) Evolution of the refractive index of the oil with temperature (triplicate).

## 6. Determination of the effective $(\mathrm{d}n/\mathrm{d}T)$ coefficient in the droplet core from comparison between experimental and numerical data

Let us consider the vertical direction that crosses the droplet along its diameter. Along the sample thickness, the integral defining the OPD can be split in two sub-integrals, one accounting for the oil layer beneath the droplet (characterized by the refractive index of oil $n_o$), and one accounting for the droplet core (characterized by the refractive index of water $n_w$):

$$\delta\ell = \frac{\mathrm{d}n_o}{\mathrm{d}T}\int_{-H}^{-2R+e}\Delta T\,dz + \frac{\mathrm{d}n_w}{\mathrm{d}T}\int_{-2R+e}^{0}\Delta T\,dz$$

This expression assumes that the $(\mathrm{d}n/\mathrm{d}T)$ coefficients of oil and water are uniform and independent of the temperature, which is commonly verified at the first order.

Reminding that, by convention, the OPD was taken in absolute values in this work, we add a conventional minus sign in the equation, and rewrite it as:

$$-\delta\ell = \alpha + \frac{\mathrm{d}n_w}{\mathrm{d}T}\beta$$

where $\alpha = \frac{\mathrm{d}n_o}{\mathrm{d}T}\int_{-H}^{-2R+e}\Delta T\,dz$, and $\beta = \int_{-2R+e}^{0}\Delta T\,dz$

The discrepancy between simulated and experimental OPD values in the droplet core (**Figure S6**a) is assumed to be due to a mismatch in the $\frac{\mathrm{d}n_w}{\mathrm{d}T}$ coefficients between simulations (where the droplet core was assimilated to pure water for which $\frac{\mathrm{d}n_w}{\mathrm{d}T} = -0.9 \times 10^{-4}\ °\mathrm{C}^{-1})^4$ and experiments (where the droplet core was loaded with polymer and gold nanoparticles). Accordingly, we may write two versions of the previous relationship, one for the experiments and one for the numerical simulations:

$$-\delta\ell_{\mathrm{exp}} = \alpha + \left(\frac{\mathrm{d}n_w}{\mathrm{d}T}\right)_{\mathrm{exp}}\beta$$

$$-\delta\ell_{\mathrm{sim}} = \alpha + \left(\frac{\mathrm{d}n_w}{\mathrm{d}T}\right)_{\mathrm{sim}}\beta$$

Considering that the simulated profiles have been optimized to fit to the experimental profiles in the oil phase, we can assume that the parameters $\alpha$ and $\beta$ are the same in the two equations.

We may eventually write an expression for the mismatch in core OPD between experiments and simulations:

$$\delta\ell_{\mathrm{exp}} - \delta\ell_{\mathrm{sim}} = -\left[\left(\frac{\mathrm{d}n_w}{\mathrm{d}T}\right)_{\mathrm{exp}} - \left(\frac{\mathrm{d}n_w}{\mathrm{d}T}\right)_{\mathrm{sim}}\right]\beta$$

$\beta$ can be directly calculated by numerical integration of the simulated temperature maps. Hence, the slope of the curve representing $\delta\ell_{\mathrm{exp}} - \delta\ell_{\mathrm{sim}}$ as a function of $\beta$ directly gives access to the difference in $\frac{\mathrm{d}n_w}{\mathrm{d}T}$ coefficients between experiments and simulations (Figure S6b).

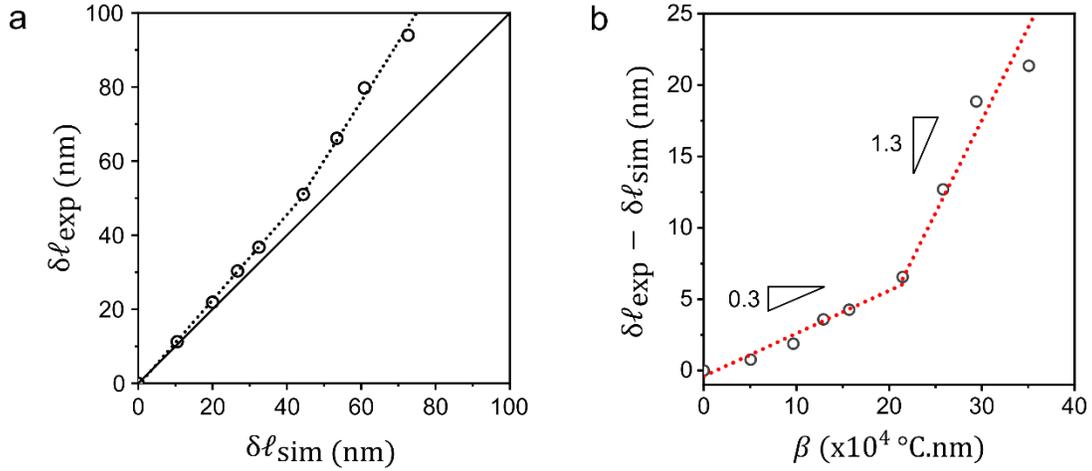

**Figure S6.** Measurement of the effective $(dn/dT)$ coefficient in the droplet core. a) Mismatch between the OPD value experimentally measured in the droplet core and the OPD value predicted by the numerical simulations (using the heating power inferred from OPD comparison in the oil phase). b) Evolution of the mismatch in core OPD (difference between experimental and numerical values) as a function of $\beta = \int_{-2R+e}^{0} \Delta T \, dz$.

Figure S6b suggests the presence of two regimes, one for temperatures lower than the LCST of PLL-g-PNIPAM (for $\beta \leq 25 \times 10^4$ °C.nm), where we have:

$$\left(\frac{\mathrm{d}n_w}{\mathrm{d}T}\right)_{\mathrm{exp}} - \left(\frac{\mathrm{d}n_w}{\mathrm{d}T}\right)_{\mathrm{sim}} \sim -0.3 \times 10^{-4} \, °\mathrm{C}^{-1},$$

and one for temperatures greater than the LCST (for $\beta > 25 \times 10^4$ °C.nm), where we have:

$$\left(\frac{\mathrm{d}n_w}{\mathrm{d}T}\right)_{\mathrm{exp}} - \left(\frac{\mathrm{d}n_w}{\mathrm{d}T}\right)_{\mathrm{sim}} \sim -1.3 \times 10^{-4} \, °\mathrm{C}^{-1}.$$

We may thus retrieve the effective values of $\frac{dn_w}{dT}$ in the droplet core: $-1.2 \times 10^{-4}$ °C$^{-1}$ before crossing the LCST, and $-2.2 \times 10^{-4}$ °C$^{-1}$ after crossing the LCST. Given that the $(dn/dT)$ of solids like AuNPs is negligible, the difference in $(dn/dT)$ with pure water can then be ascribed to the presence of polymer in its soluble state below the LCST, and to the appearance of insoluble polymer granules above the LCST.

REFERENCES RELATIVE TO THE SI

(1) Sixdenier, L.; Tribet, C.; Marie, E. Emulsion-Templated Poly(N-Isopropylacrylamide) Shells Formed by Thermo-Enhanced Interfacial Complexation. *Adv. Funct. Mater.* **2021**, *2105490* (in press). https://doi.org/10.1002/adfm.202105490.

(2) Pasparakis, G.; Tsitsilianis, C. LCST Polymers: Thermoresponsive Nanostructured Assemblies towards Bioapplications. *Polymer.* **2020**, *211*, 123146. https://doi.org/10.1016/j.polymer.2020.123146.

(3) Baffou, G. *Thermoplasmonics: Heating Metal Nanoparticles Using Light*; Cambridge University Press, 2017. https://doi.org/10.1017/9781108289801.

(4) Baffou, G. Quantitative Phase Microscopy Using Quadriwave Lateral Shearing Interferometry (QLSI): Principle, Terminology, Algorithm and Grating Shadow Description. *J. Phys. D. Appl. Phys.* **2021**, *54* (29). https://doi.org/10.1088/1361-6463/abfbf9.